\documentclass[12pt]{iopart}

\usepackage{graphicx,epsfig,floatflt}
\usepackage{amssymb}
\usepackage{iopams}
\usepackage{float}
\usepackage{multirow}

\begin{document}
\input {epsf}
\def\plotone#1{\centering \leavevmode
\epsfxsize= 0.4\columnwidth \epsfbox{#1}}
\def\plottwo#1#2{\centering \leavevmode
\epsfxsize=.43\columnwidth \epsfbox{#1} \hfil
\epsfxsize=.43\columnwidth \epsfbox{#2}}
\def\plotfiddle#1#2#3#4#5#6#7{\centering \leavevmode
\vbox to#2{\rule{0pt}{#2}}
\includegraphics{#1}}
\newdimen\hhsize\hhsize=.5\hsize
\font\lloyd=cmr8 

\def\Journal#1#2#3#4{{#1} {\bf #2}, #3 (#4)} 
\def\NCA{\em Nuovo Cimento} \def\NIM{\em Nucl. Instrum. Methods}
\def\NIMA{{\em Nucl. Instrum. Methods} A} \def\NPB{{\em Nucl. Phys.} B}
\def\PLB{{\em Phys. Lett.}  B} \def\PRL{{\em Phys. Rev. Lett.}}
\def\PRD{{\em Phys. Rev.} D} \def\ZPC{{\em Z. Phys.} C} 
\def\sst{\scriptscriptstyle} \def\mco{\multicolumn}
\def\epp{\epsilon^{\prime}} \def\vep{\varepsilon} \def\ra{\rightarrow}
\def\ppg{\pi^+\pi^-\gamma} \def\vp{{\bf p}} \def\ko{K^0}
\def\kb{\bar{K^0}} \def\al{\alpha} \def\ab{\bar{\alpha}}
\def\be{\begin{equation}} \def\ee{\end{equation}}
\def\bea{\begin{eqnarray}} \def\eea{\end{eqnarray}}
\def\CPbar{\hbox{{\rm CP}\hskip-1.80em{/}}} \def\be{\begin{equation}}
\def\ee{\end{equation}} \def\bea{\begin{eqnarray}}
\def\eea{\end{eqnarray}} \def\sm{{\rm M}_\odot}
\def\uline#1{$\underline{\smash{\hbox{#1}}}$}
\def\lta{\mathrel{\mathpalette\fun <}}
\def\gta{\mathrel{\mathpalette\fun >}}
\def\simless{\mathrel{\mathpalette\fun <}}
\def\simgreat{\mathrel{\mathpalette\fun >}} 
\def\muk{\mu{\rm K}}
\def\ang{\,{\rm\AA}}
\def\flux{\,{\rm erg\,cm^{-2}\,arcsec^{-2}\,\AA^{-1}\,s^{-1}}}
\def\GeV{\,{\rm GeV}}
\def\TeV{\,{\rm TeV}}
\def\gev{\,{\rm GeV}}
\def\keV{\,{\rm keV}}
\def\MeV{\,{\rm MeV}}
\def\sec{\,{\rm sec}}
\def\Gyr{\,{\rm Gyr}}
\def\yr{\,{\rm yr}}
\def\rcm{\,{\rm cm}}
\def\pc{\,{\rm pc}}
\def\kpc{\,{\rm kpc}}
\def\Mpc{\,{\rm Mpc}}
\def\mpc{\,{\rm Mpc}}
\def\eV{{\,\rm eV}}
\def\ev{{\,\rm eV}}
\def\erg{{\,\rm erg}}
\def\cmm2{{\,\rm cm^{-2}}}
\def\cm2{{\,{\rm cm}^2}}
\def\cmm3{{\,{\rm cm}^{-3}}}
\def\gcmm3{{\,{\rm g\,cm^{-3}}}}
\def\kms{\,{\rm km\,s^{-1}}}
\def\HO{{100h\,{\rm km\,sec^{-1}\,Mpc^{-1}}}}
\def\mpl{{m_{\rm Pl}}}
\def\mpp{{m_{\rm Pl,0}}}
\def\trh{T_{\rm RH}}
\def\g{\tilde g}
\def\R{{\cal R}}
\def\zl{z_{\rm LSS}}
\def\zeq{z_{\rm EQ}}
\def\he{$^4$He}
\def\cl{{\cal C}_l}
\def\VEV#1{\left\langle #1\right\rangle}
\def\la{\mathrel{\mathpalette\fun <}}
\def\ga{\mathrel{\mathpalette\fun >}}
\def \lleq {\lower0.9ex\hbox{ $\buildrel < \over \sim$} ~}
\def \ggeq {\lower0.9ex\hbox{ $\buildrel > \over \sim$} ~}
\def\fun#1#2{\lower3.6pt\vbox{\baselineskip0pt\lineskip.9pt
  \ialign{$\mathsurround=0pt#1\hfil##\hfil$\crcr#2\crcr\sim\crcr}}}
\def\lh{\frac{\partial\ln H}{\partial\tilde\phi}}
\def\lhh{\frac{\partial^2\ln H}{\partial\tilde\phi^2}}
\def\lhhh{\frac{\partial^3\ln H}{\partial\tilde\phi^3}}
\def\tlh{\frac{\partial\ln H}{\partial\phi}}
\def\tlhh{\partial^2\ln H/\partial\phi^2}
\def\tlhhh{\partial^3\ln H/\partial\phi^3}
\def\mpls{\frac{m^2_{\cal P}}{ 4\pi}}
\def\tmpls{ (m^2_{\cal P}/4\pi)}
\def\today{\ifcase\month\or
 January\or February\or March\or April\or May\or June\or
 July\or August\or September\or October\or November\or December\fi
 \space\number\day, \number\year}
\title{
Perturbed Power-law parameters from WMAP7 
}
\author{Minu Joy$^{a}$ and Tarun Souradeep$^{b}$ }
\address{$^a$ 
Dept. of Physics, Alphonsa College, Pala 686574, India}
\address{$^b$ 
Inter-University Centre for Astronomy and Astrophysics,
Post Bag 4, Ganeshkhind, Pune 411 007, India}



\begin{abstract}
\small{ We present a perturbative approach for studying inflation
models with soft departures from scale free spectra of the power law
model. In the perturbed power law (PPL) approach one obtains
at the leading order both the scalar and tensor power spectra with the
running of their spectral indices, in contrast to the widely used slow
roll expansion. The PPL spectrum is confronted data and we show that
the PPL parameters are well estimated from WMAP-7 data.}

\end{abstract}
\maketitle
\section{Introduction}
\label{sec:intro} 

  Observations of the anisotropy and polarization of the cosmic
microwave background (CMB) determine the parameters associated with
cosmological structure formation to unprecedented accuracy
~\cite{wmap,bond_boom,archeops}. Inflation remains the best motivated
and predictive early universe scenario that is invoked to specify the
spectrum of initial perturbations for structure formation
\cite{models}. The type of perturbation that is required results from
the simplest class of models that predict Gaussian, adiabatic, nearly
scale-invariant perturbations \cite{pert}. Generally a single
component inflaton field is considered which slowly rolls along a
sufficiently flat potential leading to the exponential expansion of
the universe. A simple and completely analytically tractable regime of
inflation corresponds to the uniform acceleration
approximation~\cite{bon_yuk} that correspond to power law inflation
model. The initial scalar and tensor perturbation spectra are scale
free (power law) and are parametrized by two numbers~: the scalar
perturbation amplitude and a common spectral index.

For the power law inflationary models~\cite{pl1,pl2}, the scale factor
$a(t)=t^p$ where $p > 1$. The slow-roll parameters for this power law
case are $\epsilon = -\delta = \frac{1}{p}$.
Also it is characterized by the (i) uniformly accelerated expansion
for which, $-\frac{\dot{H}}{H^2} = const $ and (ii) the perturbations
in the inflaton field $\delta \phi$ are equivalent to a massless
scalar field perturbations with the effective mass squared, $
m^2_{eff} = 0 $. The power law spectrum is given by $\mathcal{P}(k) =
A\,k^{(n_s-1)}$ and for this scale free case the scalar spectral index
$n_s =1 $. Here scalar and tensor perturbation spectra are of
identical shape with constant spectral indices $n_s$ and $n_t=n_s-1$;
and the ratio of amplitudes of the tensor to scalar power spectra, $r=
- 6.2 \,n_T$ \cite{liddlelyth}

It is well known that, although primordial fluctuations spectra
expected from inflation are likely to be approximately flat or
scale-invariant ($n_s(k) \equiv 1 + d\ln \mathcal{P}(k)/d\ln k \simeq
1$), exact scale invariance ($n_s=1$) is achievable only for a very
specific class of models \cite{st05}, while a slightly red spectrum
($n_s \lleq 1$) with small running appears to be a generic prediction
of the simplest viable one-parameter family of inflationary models.

In this article, we describe a perturbative formalism for studying
inflation models with soft departures from scale free spectra. The
details of our methodology to obtain the primordial power spectra is
discussed in Section \ref{sec:ppl}. And the parameter estimation with
the perturbed power law inflationary spectrum is presented in Section
\ref{sec:wmap}.

\section{Soft deviations from scale free spectra - Perturbed Power Law Model}
\label{sec:ppl} 
    Based on the `Hamilton-Jacobi formulation' \cite{salopk,lidsey},
we parametrize the inflationary phase by the Hubble parameter
$H(\phi)$, expressed as a function of the inflaton field $\phi$. For a
single inflaton field, the determination of $H(\phi)$ can be directly
translated to an estimation of the inflaton potential,
\be{}
V(\phi) = \frac{3 m^2_{\cal P}}{8\pi}
H^2(\phi) \left[1 - \frac{m^2_{\cal P}}{12\pi}\left( \tlh \right)^2 \right]
\ee{}
using the `reduced Hamilton-Jacobi equation'~\cite{bon_yuk,kinney}.

\noindent The field equation for the modes of the inflaton
perturbation, $\delta\phi_k$ are given by,
\be{}
\delta\phi_k'' + 2 \frac{a'}{a} \delta\phi_k' + (k^2 + a^2 m_{eff}^2)\delta\phi_k = 0 \ ,
\ee{}
where, $ a^2 m_{eff}^2 = \frac{a''}{a} - \frac{z''}{z}$ and $ z = a^2 \frac{\phi'}{a'}$. The effective mass can be expressed in terms of slow-roll parameters as \cite{tsthesis} ,
\be{}
\frac{m_{eff}^2}{H^2} = -(\epsilon +\delta)(\delta+3) + \frac{\dot{\epsilon}}{H} -  \frac{\dot{\delta}}{H} \ ,
\label{meff}
\ee{}
and the slow-roll parameters $\epsilon$ and $\delta$ are given by,  
\be{} \epsilon = \frac{m^2_{\cal P}}{4\pi}
\left(\frac{H_\phi}{H}\right)^2 ~~~~~~~~~~~~\mathrm{and} ~~~~~~~~~~~~
\delta = -\frac{m^2_{\cal P}}{4\pi} \frac{H_{\phi\phi}}{H} \ .
\label{epsl}
\ee{}
where the $m_{\cal P}$ is the Planck mass. In the rest of the paper
we set $\frac{m^2_{\cal P}}{4\pi} = 1$ for notational simplicity.

Starting from the equation of motion of the inflaton field and using
the Hamilton-Jacobi formalism we can show that $\frac{H_\phi}{H}=
(\tlh)$ is a constant for the power law models with uniform
acceleration. Thus it is clear from the Eqs.~(\ref{meff}) and
(\ref{epsl}) that $\frac{m_{eff}^2}{H^2}$ will be zero for the power
law models.
For the power law models, the scalar and tensor perturbation spectra
are parametrized by a common spectral index $\nu = n_s -1 = n_t$ given
by,

\be{}
\nu = \frac{-2\epsilon}{1-\epsilon} =  \frac{-2 \left(\tlh\right)^2}{\left[1- \left(\tlh\right)^2\right]} \ .
\ee{}
We are interested in considering the small deviations from uniform
acceleration in terms of a small $\tlhh$ perturbation \cite{tbond}.
In the so called Perturbed Power Law (PPL) approach, the predicted
scalar and tensor spectra are perturbed from the scale free form but,
at the leading order maintain a constant difference between their
spectral indices.  We consider that the expansion is locally modeled
to be power law with $p$ varying slowly with time. Also in terms of
the conformal time $\eta$ we can write the scale factor as $a(\eta) =
\eta^{(1/2 - \mu)}$ so that $\mu - 3/2 = 1/(p-1)$. Thus we have $\mu =
\frac{3}{2} + \frac{\epsilon(\eta)}{1 - \epsilon(\eta)} $ and also we
get,

\be{}
H^2(\phi) = \frac{(\mu - 1/2)^2}{\eta^2 a^2(\eta)} \ ,   \\
\frac{z''}{z} = \frac{1}{\eta^2} \left[\mu_S^2 - \frac{1}{4}\right] \ ,
\label{zppz}
\ee{}
\noindent where $ \mu_S^2 = \sqrt{\mu^2 - (\mu -
1/2)^2\frac{m^2_{eff}}{H^2}}$.

Here $\delta \phi$ is equivalent to the scalar field perturbations
with $ \frac{m^2_{eff}}{H^2} \simeq const $ (weakly depend on $\eta$).
In case of the PPL models, the spectral indices, $n_s \neq n_t +
1$. If we denote $n_s - 1 = \nu_s$ we can derive,
\be{} 
\nu_s = \frac{-2\epsilon(\eta)}{1-\epsilon(\eta)} +
\frac{2\chi(\eta)}{1-\epsilon(\eta)} + \mathcal{O}(\chi^2) \simeq
\frac{-2 \left[\left(\tlh\right)^2 + \left(\frac{\partial^2 \ln
H}{\partial\phi^2}\right) \right]}{[1-\left(\tlh\right)^2]} 
\ee{}
\noindent where, 
\be{}
\chi = \epsilon + \delta = - \left(\frac{\partial^2 \ln H}{\partial\phi^2}\right)\,.
\ee{}
Denoting $n_t$ as $\nu_t$ and defining the difference between the
spectral indices as, $\nu_{st} = \nu_s - \nu_t $ we get,
\be{}
\nu_{st} =  \frac{2\chi(\eta)}{1-\epsilon(\eta)} \ =  \frac{-2 \left(\frac{\partial^2 \ln H}{\partial\phi^2}\right) }{[1-\left(\tlh\right)^2]} \ ,
\ee{}
\noindent considering terms up to the first order  corrections in perturbation. It is possible to solve for $H(\phi)$ in an exact form and consequently
its $\phi$-derivatives in terms of $\nu_s$ and $\nu_{st}$. 

  By definition and using Eq.~(\ref{zppz}) the scalar perturbation
  spectrum is obtained as,
\be{}
\mathcal{P}_s(k) =  \frac{\mathcal{A}(\mu_T,\mu_S)}{2\pi} \left(\frac{H}{2\pi}\right)^2 \frac{1}{\epsilon(\eta)} 
\ee{}
and similarly the tensor perturbation spectrum, 
\be{}
\mathcal{P}_t(k) =   \frac{ 8~ \mathcal{A}(\mu_T,\mu_T)}{2\pi} \left(\frac{H}{2\pi}\right)^2 
\ee{}
\noindent where $\mathcal{A}(x,y) = \frac{4^y \Gamma^2(y)}{(x-1/2)^{2y-1}}$~, ~~$\Gamma(y)$ is the \textit{Euler Gamma Function}.
 $\mu_T$ and $\mu_S$ are defined below the Eq.~(\ref{zppz}) with the identification $\mu_T \rightarrow \mu$.  

  The standard slow-roll approximation does not consider the higher
order perturbative correction terms and thus one has $\mu_T = \mu_S =
\mu$ and there the above expressions reduces to those in
\cite{lythstewart}. It is also interesting to note that our PPL power
spectra expressions are similar to those obtained in
\cite{stewartlyth} by considering the first order corrections in
slow-roll parameters. With the PPL approach one can study both the
scalar and tensor power spectra and the running of their spectral
indices also, where as in the standard procedure one has to consider
each of them independently. While in principle we can go to
higher order PPL corrections, but we restrict up to the first order PPL
deviation from scale free spectra that the currently available data
can reliably capture.

For a given power law model, the scalar and tensor power spectra can
be calculated theoretically once we know $\nu_s $ and $\nu_{st}$. Then
it is interesting to determine how well one can constrain $H(\phi)$
using CMB observation data. The data from WMAP can be used to measure
the leading order deviations from power law spectra, quite
accurately. Parameter estimation with the perturbed power law
inflationary model is discussed in the next section.

\section{Parameter estimation with the perturbed power law model}
\label{sec:wmap} 

For Power law model, given $\nu_s $ and $\nu_{st}$ the scalar and
tensor power spectra can be calculated theoretically.  Then one has to
perform a Markov Chain Monte-Carlo sampling of the parameter space to
estimate the constraints on the inflationary parameters and the
various background cosmological parameters. We make use of the
publicly available CosmoMC package \cite{cosmomc}, which in turn uses
the CMB anisotropy code CAMB \cite{camb} to generate the theoretical
CMB angular power spectra, $C_l$s from the primordial scalar and
tensor power spectra.  We make appropriate modification to CAMB power
spectrum module to incorporate PPL power spectra.  For our analysis, we
confront the theoretical $C_l$s with the WMAP seven year data set. We
make use of the publicly available WMAP likelihood code from the
LAMBDA web site \cite{lambda}.

Flat $\Lambda$CDM is considered as the background cosmological
model. The priors that we set for the background cosmological model
are listed in \small{\tablename{~1}}. $\Omega_bh^2$ is the physical
baryon density, $\Omega_ch^2$ is the physical cold dark matter
density, $\theta$ gives the ratio of the sound horizon to the angular
diameter distance at decoupling and $\tau$ is the reionization optical
depth. The pivot point is set at $k_0 =
0.05Mpc^{-1}$. \small{\tablename{~1}} also gives the priors set for
the inflationary parameters of the PPL case, where we denote $n_s - 1
= \nu_s$ and $\nu_s - \nu_t = \nu_{st} $ and $A_s$ is the amplitude
parameter.

\begin{table}[ht]
\begin{center}
\begin{tabular}{|c|c|c|}
\hline \hline
Parameter    &  Lower Limit       	&  Upper Limit  \\
\hline
$\Omega_bh^2$   	&  0.005    		&  0.1 \\
\hline
$\Omega_ch^2$   	&  0.01      		&  0.99   \\
\hline
$\theta $        	&  0.5	  		&  10  \\
\hline
$\tau$		 	&  0.01			&  0.8 \\
\hline
$\nu_s $   	        &  -0.15    		&  0.0 \\
\hline
$\nu_{st}$   	        &  -0.06      		&  0.06  \\
\hline
$ log[10^{10} A_s] $    &  2.0	  	        &  4.2  \\
\hline \hline
\end{tabular}
\end{center}\label{tab:table2}
\caption{\footnotesize Priors for the background cosmological parameters and the PPL inflationary parameters}
\end{table}

For multiple chains the CosmoMC code computes the Gelman and Rubin
(variance of chain means)/(mean of chain variances) R statistic for
each parameter. The program also writes out the value for the worst
eigenvalue of the covariance of the means, which should be a worst
case. This $R-1$ statistic is also used for the stopping criterion
when generating chains with MPI. We set MPI Converge Stop parameter,
that can be used to stop the chains at 0.03 for our runs and got the
worst eigenvalue: $R-1 = 0.0554$.

The best fit values, the mean (of the posterior distribution of each
parameter) and 1-$\sigma$ deviations for the background parameters as
well as the PPL parameters obtained are listed in
\small{\tablename{~2}}. And for a comparative study, the similar list
for the power law (PL), power law with tensor (PLT), power law with
running (PLR), power law with running and tensor (PLRT) models are
also given in \small{\tablename{~3}}. Obviously, the best fit values
that we obtained matches with those quoted by the WMAP
team\footnote[1]{WMAP team set the pivot point at $k_0$ = 0.002
$Mpc^{-1}$.} \cite{wmap7}. It is clear from the tables that the
parameters of inflation can be well determined by the perturbative
procedure of PPL.

\begin{table}[ht]
\begin{center}
\begin{tabular}{|c|c|c|}
\hline \hline
PPL Parameter    &  Best fit        &  Mean \& 1-$\sigma$  \\
\hline
$\Omega_b~h^2$   & 0.0222     	& $0.0223^{+0.0005}_{-0.0006}$  \\
\hline
$\Omega_c~h^2$  & 0.1101       &  $0.1107^{+0.0053}_{-0.0053}$       \\
\hline
$\theta $     & 1.038 	       &   $1.039^{+0.003}_{-0.002}$  \\
\hline
$\tau$	     & 0.0881   	& $0.0876^{+0.0070}_{-0.0061}$   \\
\hline\hline\hline
\textbf{$\nu_s$}    & \textbf{-0.0367}  &  \textbf{$-0.0400^{+0.0108}_{-0.0110}$} \\
\hline
$\nu_{st}$   & \textbf{-0.0166 }     &   \textbf{$ 0.0027^{+0.0192}_{-0.0190}$}      \\
\hline\hline\hline
$ log[10^{10} A_s] $   & 3.103      &  $ 2.905^{+0.460}_{-0.368}$   \\
\hline
$A_{SZ}$         & 1.449       & $1.024^{+0.975}_{-1.024}$  \\ 
\hline
$\Omega_\Lambda$ &  0.7340 & $ 0.7320^{+0.0274}_{-0.0272}$   \\
\hline
Age/Gyr   	& 13.808     &    $ 13.784^{+0.122}_{-0.126}$      \\
\hline
$\Omega_m $     & 0.2660   &   $ 0.2680^{+0.0272}_{-0.0274}$  \\
\hline
$Z_{re} $        	& 10.595   &   $ 10.507^{+1.138}_{-1.142}$ \\
\hline
$H_0$		& 70.551   &   $ 70.663^{+2.448}_{-2.405}$\\
\hline \hline
\end{tabular}
\end{center}\label{tab:table3}
\caption{\footnotesize The best fit values, the mean and 1-$\sigma$ deviations for the various input and derived parameters obtained by comparing PPL spectrum with WMAP7 data}
\end{table}
\begin{table}[!hbt]
\begin{center}
\begin{scriptsize}
\begin{tabular}{|c|c|c|c|c|c|c|c|c|}
\hline \hline
Model	& \multicolumn{2}{c|}{PL} & \multicolumn{2}{c|}{PLT} & \multicolumn{2}{c|}{PLR} & \multicolumn{2}{c|}{PLRT} 
\\
\cline{1-9}

Parameter & Best fit & Mean \& 1-$\sigma$ & Best fit  & Mean \& 1-$\sigma$ & Best fit  & Mean \& 1-$\sigma$ & Best fit  & Mean \& 1-$\sigma$ \\
\hline
$\Omega_b~h^2$   	
& 0.0221 &  $0.0223^{+0.0006}_{-0.0005}$ 
& 0.0222 &  $0.0223^{+0.0006}_{-0.0005}$ 
& 0.0213 &  $0.0216^{+0.0008}_{-0.0007}$ 
& 0.0214 &  $0.0216^{+0.0008}_{-0.0007}$\\
\hline
$\Omega_c~h^2$ 
& 0.1090 &  $0.1106^{+0.0054}_{-0.0054}$ 
& 0.1116  & $0.1106^{+0.0053}_{-0.0053}$ 
& 0.1202 & $0.1180^{+0.0073}_{-0.0076}$ 
& 0.1175 & $0.1184^{+0.0083}_{-0.0085} $\\  
\hline
$\theta $ 
&1.038 & $1.039^{+0.003}_{-0.002}$
&1.038 & $1.039^{+0.003}_{-0.002}$
&1.037 & $1.038^{+0.003}_{-0.002}$
&1.038 & $1.038^{+0.003}_{-0.003}$  \\
\hline
$\tau$
&0.0840 & $0.0867^{+0.0071}_{-0.0066}$
&0.0829 & $0.0864^{+0.0072}_{-0.0061}$
&0.0900 & $0.0914^{+0.0077}_{-0.0067}$
&0.0961 & $0.0914^{+0.0080}_{-0.0069} $  \\
\hline\hline\hline
$\textbf{n}_\textbf{s} $
&0.9556 & $0.9628^{+0.0129}_{-0.0204}$
&0.9608 & $0.9620^{+0.0136}_{-0.0133}$
&0.9034 & $0.9175^{+0.0367}_{-0.0358}$
&0.9108 & $0.9146^{+0.0383}_{-0.0386}$   \\
\hline
$\textbf{n}_{\textbf{t}}$
& NA &
&0.3881 & $-0.0082^{+0.3286}_{-0.3232}$
& NA & 
& -0.2321 & $0.0047^{+0.3322}_{-0.3324}$   \\
\hline
$\textbf{n}_{\textbf{run}}$
& NA &
& NA &
& -0.0387 & $-0.0328^{+0.0238}_{-0.0226}$
& -0.0367 & $-0.0351^{+0.0265}_{-0.0265}$   \\
\hline\hline\hline
$r$
& NA &
&0.0175  & $0.1134^{+0.1134}_{-0.0265}$
& NA &
& 0.0177 & $0.1300^{+0.1300}_{-0.0343}$   \\
\hline
$ log[10^{10} A_s] $
&3.050	& $ 3.071^{+0.035}_{-0.055}$
&3.069  & $ 3.072^{+0.037}_{-0.033}$
&3.098  & $ 3.097^{+0.039}_{-0.040}$
&3.091  & $ 3.099^{+0.042}_{-0.042}$   \\
\hline
$A_{SZ}$
&1.935& $1.058^{+0.942}_{-1.058}$
&1.025& $1.068^{+0.932}_{-1.068}$
&1.411& $1.059^{+0.941}_{1.059}$
&1.352& $1.035^{+0.965}_{1.035}$ \\
\hline
$\Omega_\Lambda$
&0.7381	& $0.7325^{+0.0278}_{-0.0272}$ 
&0.7270 & $0.7323^{+0.0285}_{-0.0273}$
&0.6724 & $0.6857^{+0.0472}_{-0.0456}$
&0.6910 & $0.6825^{+0.0526}_{-0.0522}$ \\
\hline
Age/Gyr
&13.823	& $ 13.785^{+0.121}_{-0.126}$
&13.814 & $ 13.788^{+0.126}_{-0.131}$
&13.984 & $ 13.921^{+0.160}_{-0.155}$
&13.939 & $ 13.929^{+0.166}_{-0.163}$
\\
\hline
$\Omega_m $
&0.2619  & $0.2675^{+0.0715}_{-0.0744}$
&0.2730  & $0.2677^{+0.0273}_{-0.0284}$
&0.3276  & $0.3143^{+0.0456}_{-0.0473}$
&0.3090  & $0.3175^{+0.0522}_{-0.0526}$   \\
\hline
$Z_re $
&10.248 & $10.428^{+1.161}_{-1.162}$
&10.197 & $10.408^{+1.180}_{-1.176}$
&11.317 & $11.239^{+1.442}_{-1.405}$
&11.266 & $11.286^{1.395}_{-1.436}$    \\
\hline
$H_0$
&70.775	& $70.694^{+2.375}_{-2.406}$
&70.020 & $70.663^{+2.492}_{-2.386}$
& 65.740 & $67.083^{+3.468}_{-3.308}$
& 67.043 & $ 66.903^{+3.741}_{-3.844}$   \\
\hline \hline
\end{tabular}
\end{scriptsize}
\end{center}\label{tab:table4}
\caption{\footnotesize The best fit values, the mean and 1-$\sigma$ deviations for the various input
and derived parameters obtained by comparing power law (PL), power law with tensor (PLT), power law 
with running (PLR), power law with running and tensor (PLRT) models spectra with WMAP7 data}
\end{table}

The least square likelihood parameter $\chi_{eff}^{2}$ for 
the perturbed power law (PPL), power law (PL), power law with tensor (PLT), power law 
with tensor obeying the consistency relation $ n_t = - r/ 6.2 $ (PLTC), 
power law with running of scalar spectral index (PLR), 
power law with running and tensor (PLRT) and the 
power law with running and tensor obeying the consistency relation (PLRTC) models
are given in \tablename{~4}.
The first raw gives the $\chi_{eff}^{2}$
value obtained by fitting the power spectrum obtained for a power law
model with the perturbation approach (PPL), to the WMAP7 spectrum.  We can
see that the values of first four runs are almost the same (the best
one being that of PPL), where as we get a slightly better
$\chi_{eff}^{2}$ for a model with the running of the spectral index
that involve more parameters. The second column of the table gives the number of extra inflationary 
parameters compared to the simple power law model. For PL the inflationary paramters are the scalar spectral index $n_s$ and 
the scalar power amplitude $A_s$. To obtain the tensor power spectrum (PLT) we need two more paramters, 
the tensor spectral index $n_t$ and the ratio of scalar and tensor power amplitudes $r = A_T /A_S $. For PLTC model, 
we impose the consistency relation $n_t = - r/6.2 $ also. For PLR model the running of scalar spectral index $n_{run}$ is also 
considered. Similarly, the extra paramters of the PLRT model are $n_t$, $r$ and $n_{run}$. And $r$ and $n_{run}$ for the PLRTC model 
where the consitency relation is also imposed.  It is interesting to note that in the case of PPL just one extra parameter $\nu_{st}$
capture the role of $n_t$, $n_{run}$ and in addition the running of tensor spectral index also, which is not possible to be estimated by the standard
PL spectra as it is too small. 
\begin{table}[!htb]
\begin{center}
\begin{footnotesize}
\begin{tabular}{|c|c|c|}
\hline\hline
Model					&  No. of extra		& $\chi_{eff}^{2}$ \\
					& Parameters 		&		\\
\hline
\textbf{PPL}		  		& 1			& $7474.96$\\
\hline
PL				        & 0			& $7475.14$\\
\hline
PLT 			                & 2	    	  	& $7475.00$\\
\hline
PLTC 					& 1 			& $7475.07$\\ 
\hline
PLR 					& 1			& $7473.74$\\
\hline
PLRT					& 3 			& $7473.58$\\
\hline
PLRTC 					& 2 			&  $7473.61$\\
\hline\hline
\end{tabular}
\caption{\label{tab:chsq}The best fit likelihood $\chi_{eff}^{2}$ values for the perturbed power law (PPL), power law (PL), power law with tensor (PLT), power law 
with tensor obeying the consistency relation $ n_t = - r/ 6.2 $ (PLTC), power law with running of scalar spectral index (PLR), power law with running and tensor (PLRT),
power law with running and tensor obeying the consistency relation (PLRTC) models.
}
\end{footnotesize}
\end{center}\label{tab:table5}
\end{table}

1D marginalized posterior distribution of the background cosmological
 parameters obtained by PPL fitting to WMAP7 data is plotted in
 \figurename{~1}. It follows that there are no significant changes in
 derived values of the cosmological parameters in comparison with the
 results obtained by the WMAP team assuming a power-law model of the
 primordial spectrum.

\begin{figure}[!ht]
\begin{center}
\resizebox{340pt}{320pt}{\includegraphics{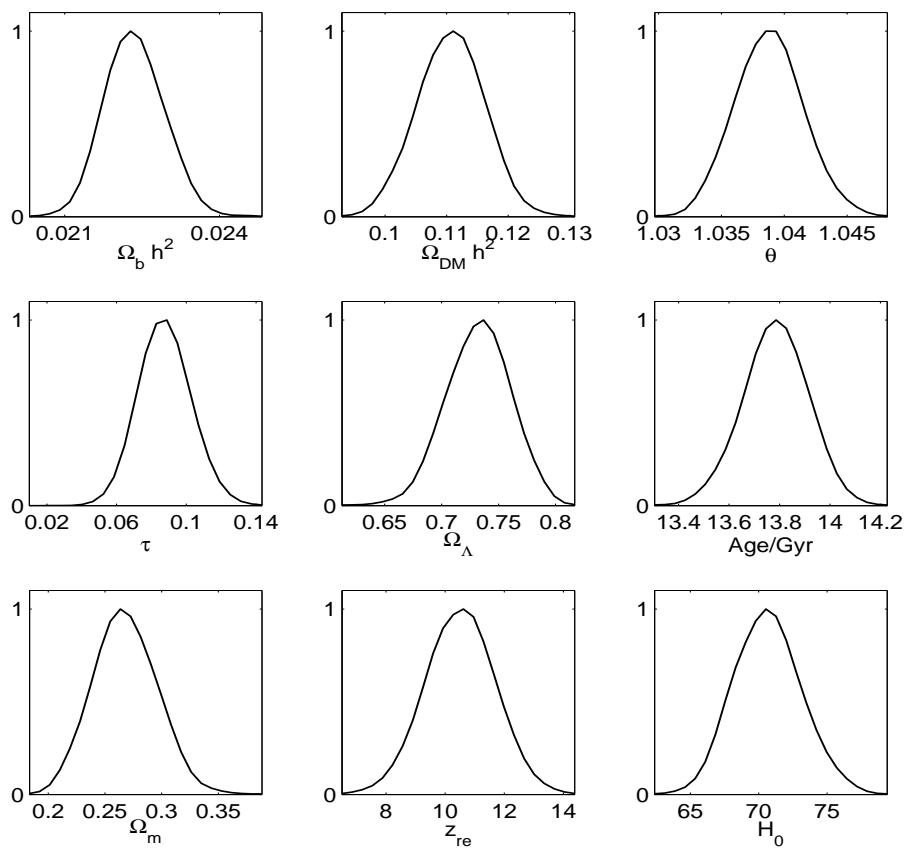}}
\label{fig:figure1}
\end{center}
\caption{\small 1D marginalized posterior distribution of the
 background cosmological parameters (both input and derived) obtained
 by PPL fitting to WMAP7 data. Y-axis gives the probability from 0 to
 1.  }
\end{figure}

\figurename{~2} gives the 1D marginalized posterior distribution of
the PPL inflationary parameters. The marginalized probability for
$\nu_s$ peaks at $-0.04$ (corresponding to $n_s = 0.96$ at the pivot
point), and the marginalized probability for $\nu_{st}$ is found to be
maximum at 0.0027. The $95\%$ marginalised limit of $\nu_{st}$ is found to be in the
range $-0.016 < \nu_{st} < 0.022 $. The plot of joint 2D marginalized
posterior distribution of $\nu_s$ and $\nu_{st}$ is given by
\figurename{~3}.  The colour of the figure shows how many times the
CosmoMC has probed a specific area of the parameter space and the
inner and outer closed contour lines indicate the 1 and 2-$\sigma$
likelihood contours.


\begin{figure}[!hbt]
\begin{center}
\resizebox{220pt}{320pt}{\includegraphics{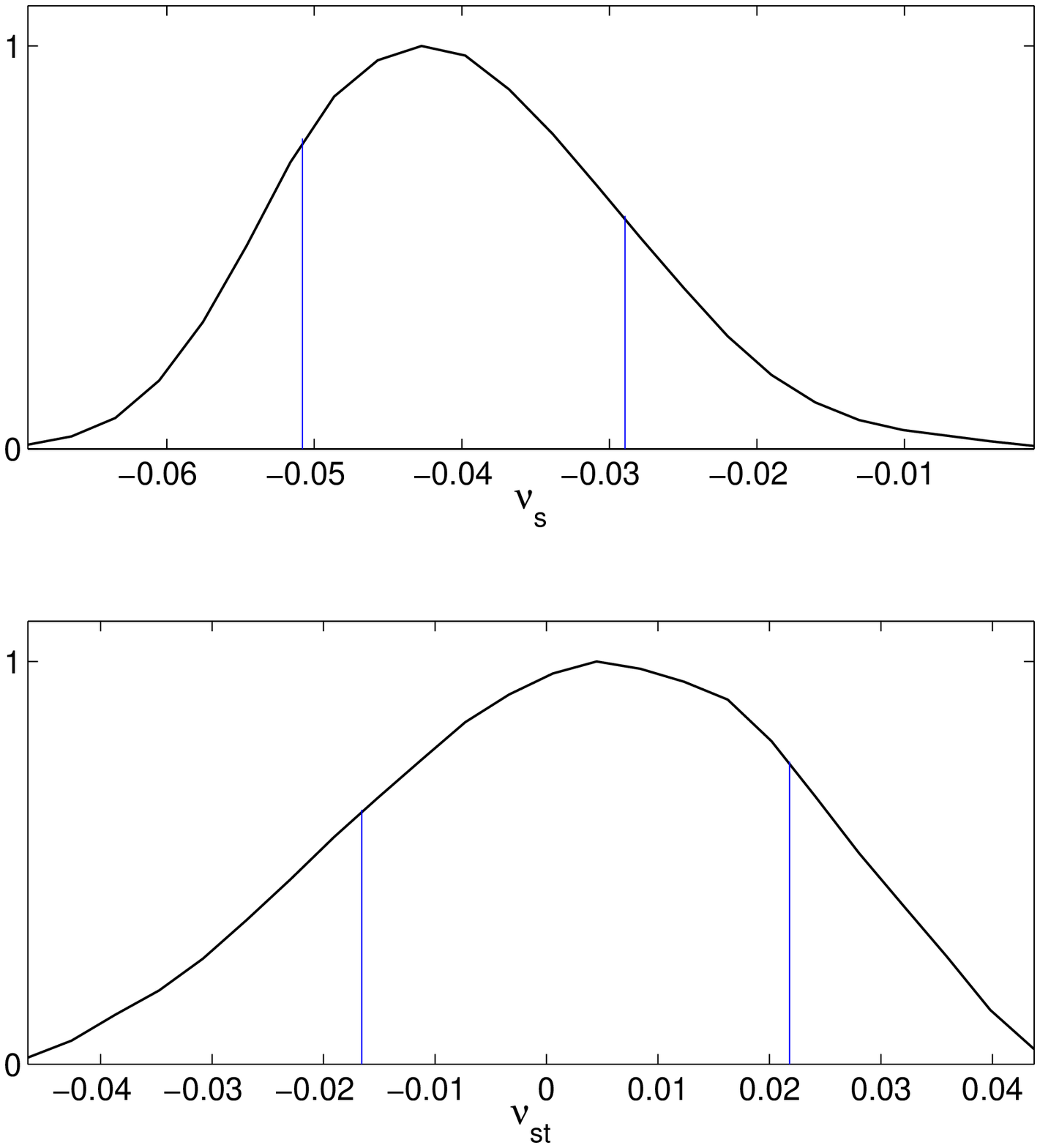}}
 \caption{\small 1D marginalized posterior distribution of the PPL
 inflationary parameters $\nu_s$ and $\nu_{st}$ obtained by fitting to
 WMAP7 data. The $95\%$ marginalised limit on the parameters are shown by the vertical lines.}
\label{fig:figure2}
\end{center}
 \end{figure}


\begin{figure}[!hbt]
\begin{center}
\resizebox{220pt}{200pt}{\includegraphics{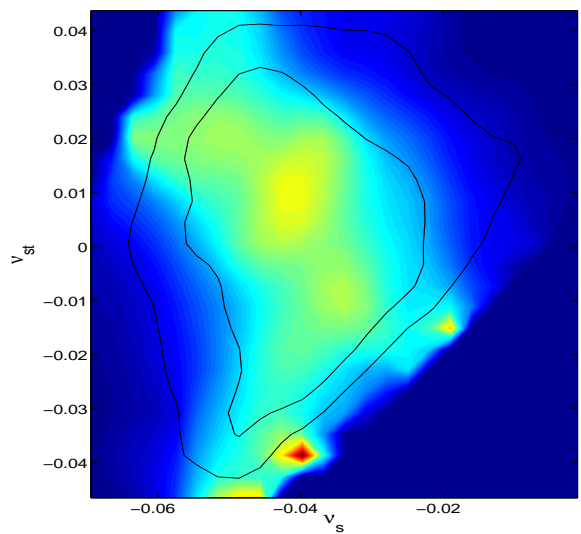}}
\caption{\small
The joint 2D marginalized posterior distribution of  $\nu_s$ and $\nu_{st}$
}
\label{fig:figure3}
\end{center}
\end{figure}


\begin{figure}[tbh!]
\begin{center}
\resizebox{320pt}{240pt}{\includegraphics{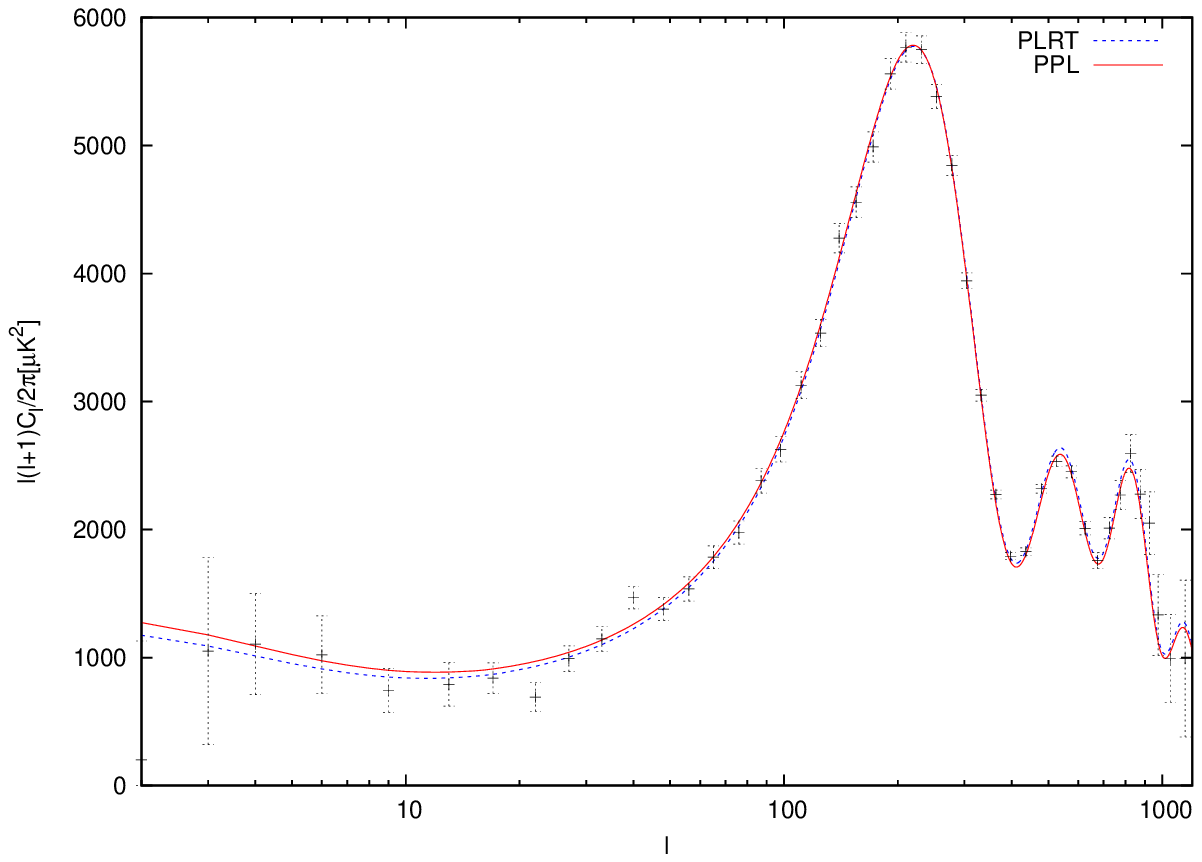}}
\caption{\small
Best fit $C_{\ell}s$ for power-law (PLRT)~(blue, dotted), power law with perturbative method (PPL) ~(red, solid) and the WMAP7 binned data with related error bars }
\end{center}
\label{fig:Cls}
\end{figure}

The best fit $C_{\ell}s$ for power law with running and tensor (PLRT)
 and power law with perturbative method (PPL) are presented in
 \figurename{~4}.  The WMAP7 binned data with related error bars are
 also plotted for comparison. It is clear that the PPL spectrum gives
 a very good fit to the observed data.

\section{Conclusions and discussion}

The high quality CMB data that has become available over the past few
 years and other observational data are just approaching the level of
 accuracy necessary to detect deviations from exact scale invariance
 and to distinguish between different inflationary models. The data
 indicates that the departure of the spectral index from exact scale
 invariance is likely to be small, $|n_s(k)-1|\ll 1$, which is in good
 agreement with predictions of the simplest inflationary scenarios. A
 perturbative procedure for studying inflation models with soft
 departures from scale free spectra is discussed in this paper.  In
 \cite{tbond} one of the authors forecast how well one can constrain
 $H(\phi)$ using WMAP and Planck data. The expected 1-$\sigma$ error
 ellipses \cite{tbond} in the $(\partial \ln H/ \partial \phi)^2$ - $(\partial ^2 \ln
 H/ \partial \phi^2)$ plane for a fixed target power law model, are
 now obtained with actual WMAP data as given in \figurename{~3}.

The perturbed power law spectrum is confronted with the 7-year WMAP
 data. The best fit values that we obtained for both the input and
 derived parameters matches very well with those quoted by the WMAP
 team for power law inflationary models. Also it is interesting to
 note that the $\chi_{\rm eff}^{2}$ value we obtained by fitting the
 PPL spectrum to the WMAP7 spectrum is very slightly better than that
 for a simple power law spectrum (with and without tensors). We find
 that the parameters of inflation can be well determined by this
 perturbative method. In case of the standard power law spectrum one
 has to consider the scalar and tensor spectral indices $n_s$, $n_t$
 and the running of scalar spectral index $n_{run}$ as separate
 independent parameters. Here, in the case of PPL just the two
 parameters $\nu_s$ and $\nu_{st}$ will capture all these. In
 addition, PPL spectra takes care of the running of tensor spectral
 index also, which is not possible to be estimated by the standard PL
 spectra as it is too small. In conclusion, the perturbed power law
 model helps to determine how the general models of inflation are
 best studied as `departures from the power law model'. The probable
 values of the `deviation parameter' $\nu_{st}$ is found to be
 consistent with zero (contained between $-0.04 < \nu_{st} <
 0.04$). More precise data from Planck \cite{planck} would be able to
 measure the deviations from power law spectra quite accurately and
 our PPL approach would help to analyse and understand the results in
 a better way.

\section{Acknowledgments}
We acknowledge the use of high performance computing system at
 IUCAA. We thank Rita Sinha for contributions during the early stages
 of this project. We thank Moumita Aich for useful
 commments on our paper. MJ acknowledges the support from U.G.C. Major
 research project grant, F.No.34-30/2008(SR) and the Associateship of
 IUCAA.  TS acknowledges support from Swarnajayanti Fellowship, DST,
 India.



\end{document}